\documentclass[master]{ws-rv9x6} 
\usepackage{ws-rv-van}    
\usepackage{ws-rv-thm}    
\usepackage{subfigure}    
\usepackage{ws-index}     

\makeindex
\newindex{aindx}{adx}{and}{Author Index}       
\renewindex{default}{idx}{ind}{Subject Index}  

\begin{document}

\chapter{Reaction kinetics in the few-encounter limit}

\author[D. Hartich and A. Godec]{David Hartich and Alja\v{z} Godec}
\index[aindx]{Hartich, D.} 
\index[aindx]{Godec, A.}

\address{Mathematical Biophysics Group, Max-Planck-Institute for Biophysical Chemistry, 37077 Göttingen, Germany\\
david.hartich@mpibpc.mpg.de\\
agodec@mpibpc.mpg.de
}

\begin{abstract}
The classical theory of chemical reactions can be understood
in terms of diffusive barrier crossing, where
the rate of a reaction is determined by the inverse of the mean first
passage time (FPT) to cross a free energy
barrier. Whenever a few reaction events suffice to trigger a response or the energy barriers are not high, the mean first
passage time alone does not suffice to characterize the kinetics,
i.e., the kinetics do not occur on a single time-scale. Instead, the
full statistics of the FPT are required. 
We present a spectral representation of the FPT statistics
that allows us to understand and accurately determine FPT distributions
over several orders of magnitudes in time.
A canonical narrowing of the first passage density is shown to emerge 
whenever several molecules are searching for the same target, which
was termed the \emph{few-encounter limit}.
The few-encounter limit is essential in all situations, in which
already 
the first encounter triggers a response, such as misfolding-triggered aggregation of proteins or protein transcription regulation.
\end{abstract}

\section{Introduction}
Since Smoluchowski's \cite{smol16} and Kramers' \cite{kram40} seminal contributions
first passage time (FPT) theory has been a paradigm for studying chemical
kinetics \cite{szab80,ben93,osha95,guer16,li17}, see also
Refs.~\refcite{haen90,redn01,bray13,metz14a,beni14} for extensive reviews.
Extensions of these original ideas led to theories of diffusion-controlled reaction
kinetics in fractal \cite{kope88,avra00} and heterogeneous media \cite{bres13,gode15,vacc15,gode16a},
surface-mediated reactions \cite{rupp15,greb16}, and search processes
involving swarms of agents \cite{meji11}, to name but a
few. Notably, in contrast to extensively studied nearest-neighbor
random walks (see e.g. Ref.~\refcite{beni14}), the FPT statistics in multiply-connected
Markov-state dynamics, aside from a few studies on simple enzyme models
\cite{muns09,bel10,grim17} and recent numerical approximation schemes based on
Bayesian inference \cite{schn17,webe17}, are barely explored
\cite{schn17a}.

The importance of understanding the full FPT statistics is meanwhile
well established
\cite{beni10,meye11,beni14,meji11,gode16}. For example, it was
proposed to be essential for explaining the so-called proximity effect
in gene regulation, according to which direct reactive trajectories
boost the speed and precision of gene regulation
\cite{kole07,fras07}. The full FPT statistics were also shown to
be required for
a quantitative description of misfolding-triggered protein aggregation  \cite{hart18_arxiv}
and various nucleation-limited phenomena
\cite{kamp93,beij03}. Underlying the kinetics in these systems is the
FPT problem of $n$-independent simultaneous trajectories
\cite{gode16,hart18_arxiv}, which we refer to as kinetics in the
\emph{few encounter limit} and will be the focus of this chapter.

We will limit the discussion to FPT phenomena of reversible Markovian
dynamics in bounded domains or
confining potentials, which renders all FPT moments finite and probability densities asymptotically
exponential \cite{beni10,meye11,beni14,meji11,gode16,hart18_arxiv}. We
discuss effectively one-dimensional diffusion processes in arbitrary
potentials $U(x)$ and jump processes with arbitrary transition
matrices. Hyperspherically symmetric
diffusion processes in $d$ dimensions will be treated via a mapping onto radial
diffusion with a repulsive potential $U(x)=(d-1)\ln(x)$
in units of thermal energy \cite{redn01,beni10,meye11,beni14,gode16},
i.e., $k_\mathrm{B}T\equiv 1$.

\begin{figure}[t]
\centering
\includegraphics[width=0.9\textwidth]{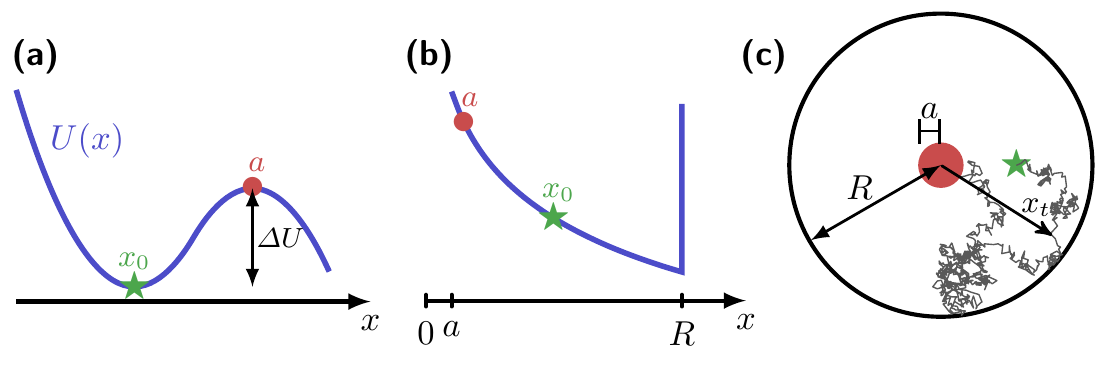}
\caption{
 Crossing of a free energy barrier. (a) Energy landscape $U(x)$ from a
 local potential minimum $a$ over a barrier $a$ with barrier height
 $\varDelta U$. (b) Geometry-induced potential $U(x)=-(d-1)\ln x$ for a
 diffusive search in a $d$-dimensional domain with radius $R$ as
 illustrated in (c).}
 \label{fig:HartichGodec:ill_intro}
\end{figure}

The chapter is organized as follows.
Generic single-molecule FPT concepts are introduced in
Sec.~\ref{sec:HartichGodec:single_molecule}. Section
\ref{sec:HartichGodec:single_molecule_spectral} outlines a spectral
expansion of the FPT density. Section
\ref{sec:HartichGodec:multi-molecule} relates the single-molecule
FPT problem to the corresponding many-particle problem, while two examples
of FPT statistics in discrete- and continuous state-space
dynamics are presented in
Sec.~\ref{sec:HartichGodec:defining_kinetics}. In working out these
examples we utilize a recently proven duality between first passage
and relaxation processes -- an algorithmic tool that allows determining the full
first passage time distribution analytically from a simpler relaxation
process (Appendix \ref{sec:HartichGodec:duality}). We conclude with an
outlook in Sec.~\ref{sec:HartichGodec:conclusion}.

\section{First passage time statistics}
\label{sec:HartichGodec:FPTdistribution}
\subsection{The single-particle setting}
\label{sec:HartichGodec:single_molecule}
Let $x_t$ denote the dynamics of a reaction coordinate, e.g., the position of
a particle in a potential $U(x)$ (see
Fig.~\ref{fig:HartichGodec:ill_intro}a) or in a circular domain with a
central target (see
Fig.~\ref{fig:HartichGodec:ill_intro}b,c). Suppose that $x_t$
obeys a Markovian equation of motion. The reaction kinetics
are then characterized by the FPT -- the first instance $x_t$ reaches
a given threshold $a$, defined formally as
\begin{equation}
 t_a(x_0)=\min\{t|x_t=a\}.
 \label{eq:HartichGodec:def_FPT}
\end{equation}
The stochasticity of $x_t$
renders the FPT, $t_a(x_0)$, a stochastic variable. The statistics
of $t_a(x_0)$ is fully characterized by the survival 
probability
\begin{equation}
 \mathcal{P}_a(t|x_0)\equiv \operatorname{Prob}[t_a(x_0)\ge t],
 \label{eq:HartichGodec:Pt}
\end{equation}
which quantifies the probability that the reaction did not occur before $t$. 
$\mathcal{P}_a(t|x_0)$ decays monotonically from
$\mathcal{P}_a(t|x_0)=1$ to $\mathcal{P}_a(t{=}\infty|x_0)=0$
with a slope that
is nothing but the first passage time density
\begin{equation}
 \wp_a(t|x_0)\equiv-\frac{\partial}{\partial t}\mathcal{P}_a(t|x_0).
 \label{eq:HartichGodec:pt}
\end{equation}
The $k$th moment of $t_a(x_0)$
can be determined via
\begin{equation}
 \langle t_a(x_0)^k\rangle=\int_0^\infty t^k \wp_a(t|x_0)\mathrm{d}t
 =k\int_0^\infty t^{k-1} \mathcal{P}_a(t|x_0)\mathrm{d}t,
 \label{eq:HartichGodec:moments}
\end{equation}
where the last equality follows from Eq.~\eqref{eq:HartichGodec:pt} by
partial integration. Notably,  $\langle t_a(x_0)^k\rangle$ are typically dominated by the
long-time behavior of $\wp_a(t|x_0)$ \cite{hart18_arxiv,gode16}.
While the full FPT density is generally hard to determine, simple integral
formulas exist for the moments of the FPT under diffusive dynamics
\cite{gard04}.

However, as we show later, the moments of the FPT in the
single-particle setting in fact provide very little information about the
kinetics in many-particle systems. Namely, few-encounter and
nucleation kinetics for example, are typically governed by short \cite{kamp93,beij03} or intermediate time-scales \cite{hart18_arxiv}.


\subsection{Spectral expansion of first passage distributions}
\label{sec:HartichGodec:single_molecule_spectral}
For reversible Markovian dynamics the FPT density allows the expansion
\begin{equation}
\wp_a(t|x_0)=\sum_{k>0} w_k(x_0)\mu_k\mathrm{e}^{-\mu_k t},
\label{eq:HartichGodec:spectral_def}
\end{equation}
where $\mu_k^{-1}$ denotes the $k$th first passage time-scale such that $\mu_k$ is a rate,
and $w_k(x_0)$ is the corresponding weight of the $k$th mode. In contrast to $\mu_k$, $w_k(x_0)$ depends on the starting position $x_0$. For convenience, we drop the functional dependence of both $w_k$ and $\mu_k$ on $a$. 
The weights satisfy the normalization condition $\sum_k w_k(x_0)=1$, and
the positivity of $\wp_a(t|x_0)$ implies $w_1>0$. Specifically, if
energetic or kinetic barriers are high enough a separation of
time-scales emerges ($\mu_2\gg\mu_1$), such that the FPT distribution
becomes approximately $\wp_a(t|x_0)\simeq w_1(x_0)\mu_1\mathrm{e}^{-\mu_1 t}$ with $w_1(x_0)\simeq 1$ if $x_0$ is located \emph{before} the highest energy barrier \cite{gode16}.
Note that for finite discrete-state systems the sum in Eq.~\eqref{eq:HartichGodec:spectral_def} is finite.
The survival probability analogously becomes
\begin{equation}
 \mathcal{P}_a(t|x_0)=\int_t^\infty\wp_a(t'|x_0)\mathrm{d}t'=\sum_{k>0} w_k(x_0)\mathrm{e}^{-\mu_k t}.
\end{equation}
Using the spectral expansion the Laplace transform of $\wp_a(t|x_0)$
reads 
\begin{equation}
\tilde\wp_a(s|x_0)=\int_0^\infty\mathrm{e}^{-st}\wp_a(t|x_0)\mathrm{d}t=\sum_{k>0} \frac{w_k(x_0)\mu_k}{s+\mu_k}.
\label{eq:HartichGodec:FPD_Laplace}
\end{equation}
and the $k$th moment of the single-particle first passage time is given by
\begin{equation}
 \langle t_a(x_0)^k\rangle=k!\sum_{i>0}w_i(x_0)\mu_i^{-k}.
 \label{eq:HartichGodec:moments_spectral}
\end{equation}
When  $\mu_2\gg\mu_1$, $\langle t_a(x_0)^k\rangle$ is typically dominated by the slowest
time scale, i.e., $\langle t_a(x_0)^k\rangle\simeq k!w_1(x_0)/\mu_1^k$, which is usually quite accurate in problems such as the one used in Fig.~\ref{fig:HartichGodec:ill_intro} (see also Ref.~\refcite{hart18_arxiv}).

In general it can be difficult to determine both, first passage eigenvalues $\{\mu_k(x)\}$
and their corresponding weights $\{w_k(x)\}$. However, we have
recently derived an analytical theory that
allows us to determine the spectral representation of $\wp_a(t|x_0)$
from the corresponding dual relaxation spectrum
\cite{hart18_arxiv,hart18a_arxiv}, which is summarized in Appendix~\ref{sec:HartichGodec:duality}.

\subsection{The many-particle setting and kinetics in the few-encounter limit}
\label{sec:HartichGodec:multi-molecule}
Suppose that now $n$ particles starting from the same position $x_0$ at time $t=0$
are searching independently for the same target. Once the first molecule hits the target
a ``catastrophic'' response is triggered (e.g., aggregation of
misfolded of proteins, induction/inhibition of gene transcription
etc.), or the target disappears such as in foraging problems. In order to understand such ``nucleation-type phenomena'' details about the first passage time distribution become relevant \cite{beni10,beni14,gode16,greb18}.
The $n$-particle survival probability
is simply the product of the single particle survival 
probabilities \cite{meji11}
\begin{equation}
 \mathcal{P}_a^{(n)}(t|x_0)\equiv \mathcal{P}_a(t|x_0)^n.
 \label{eq:HartichGodec:Ptn}
\end{equation}
The probability density that the first of the $n$ particles hits $a$ for the first time at time $t$ then
becomes using  Eqs.~\eqref{eq:HartichGodec:pt} and
\eqref{eq:HartichGodec:Ptn} \cite{meji11,hart18_arxiv,hart18a_arxiv}
\begin{equation}
\!\!\wp_a^{(n)}(t|x_0)\equiv  n\wp_a(t|x_0)\mathcal{P}_a(t|x_0)^{n-1}
  =n\wp_a^{(n)}(t|x_0)\left[\int_t^\infty\!\!\wp_a(\tau|x_0)\mathrm{d}\tau\right]^{n-1}.
  \label{eq:HartichGodec:ptn}
\end{equation}
We will henceforth omit the superscript $(1)$ in denoting the single-particle scenario, i.e., $\wp_a\equiv\wp_a^{(1)}$ and  $\langle\cdots\rangle\equiv \langle\cdots\rangle^{(1)}$.
\begin{figure}[t]
 \centering
 \includegraphics{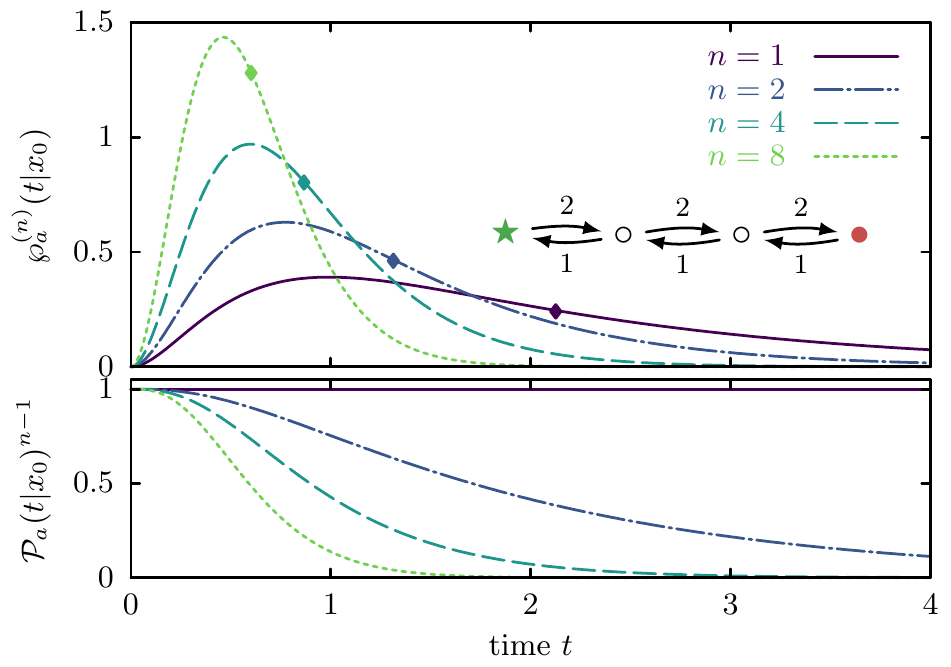}
 \caption{$n$-particle density
   $\wp_a^{(n)}(t|x_0)$ (top) and the second term in the product
   of  Eq.~(\ref{eq:HartichGodec:ptn}), $\mathcal{P}_a(t|x_0)^{n-1}$, (bottom)
   for a four-state random walk. Diamonds depict the
   respective mean FPTs. 
   More details about $\wp_a$ and $\mathcal{P}_a$ are given in Sec.~\ref{sec:HartichGodec:example_discrete}.}
 \label{fig:HartichGodec:pt_ill}
\end{figure}
Analogously to Eq.~\eqref{eq:HartichGodec:moments},
the moments of the first passage time in the  $n$-particle case
read
\begin{equation}
 \langle t_a(x_0)^k\rangle^{(n)}\equiv \int_0^\infty t^k\wp_a^{(n)}(t|x_0)\mathrm{d}t,
 \label{eq:HartichGodec:moments_n}
\end{equation}
which according to Eq.~\eqref{eq:HartichGodec:ptn}
can be determined solely from $\wp_a(t|x_0)$
\begin{align}
  \langle t_a(x_0)^k\rangle^{(n)}&=
  n\langle t_a(x_0)^k\mathcal{P}_a[t_a(x_0)|x_0]^{n-1}\rangle\nonumber\\
  &=n\int_0^\infty t^k\mathcal{P}_a(t|x_0)^{n-1}\wp_a(t|x_0)\mathrm{d} t.
  \label{eq:HartichGodec:mean_n_vs_1}
\end{align}
Due to the term $\mathcal{P}_a(t|x_0)^{n-1}$ in Eq.~\eqref{eq:HartichGodec:mean_n_vs_1} one needs, for any finite value of $k\ge1$,
formally an infinite number of single-particle moments, to
determine $\langle t_a(x_0)^k\rangle^{(n)}$. Hence, many-particle nucleation-type kinetics cannot be understood in terms of single-particle mean first passage times \cite{kamp93,beij03,hart18_arxiv}.

Even if $\langle t_a(x_0)^k\rangle$ is accurately characterized by long-time asymptotics, the latter do not provide accurate results for $\langle t_a(x_0)^k\rangle^{(n)}$, which can be orders of magnitude off \cite{hart18_arxiv}. The severe insufficiency of single-particle moments arises
from the sharp sigmoidal shape of $\mathcal{P}_a(t|x_0)^{n-1}$ within the many-particle average
\eqref{eq:HartichGodec:mean_n_vs_1} (e.g., see lower panel of Fig.~\ref{fig:HartichGodec:pt_ill} for an illustration).
We note that utilizing long-time asymptotics can lead in general to both an overestimation or an underestimation
of $\langle t_a(x_0)^k\rangle^{(n)}$, depending on the initial conditions \cite{hart18_arxiv}.
For $n\to\infty$ short-time asymptotics sets in, 
for which it has been found that
$\langle t_a(x_0)\rangle^{(n)}\propto1/\ln(n)$ for overdamped diffusive
first passage problems \cite{kamp93,beij03} (see also Ref.~\refcite{meji11}).

Two generic phenomena emerge as
the particle number $n$ increases: (i)  $\langle
t_a(x_0)\rangle^{(n)}$ reduces and (ii) the width of
$\wp_a^{(n)}(t|x_0)$ concurrently decreases (see
Fig.~\ref{fig:HartichGodec:pt_ill}). Both are independent of the
details of dynamics and directly follow from a progressively sigmoidal
shape of $\mathcal{P}_a(t|x_0)^{n-1}$. These features are particularly
important for explaining the so-called proximity effect -- the
spatial proximity of co-regulated genes -- in
transcription regulation \cite{gode16}.
The generic origin of these effects provides an explanation of the robustness of the proximity
effect (see  Ref.~\refcite{gode16} for more details
on the biological aspect).

\subsection{Determining first passage time statistics from
  relaxation spectra}
\label{sec:HartichGodec:defining_kinetics}

\begin{figure}
\centering
\includegraphics{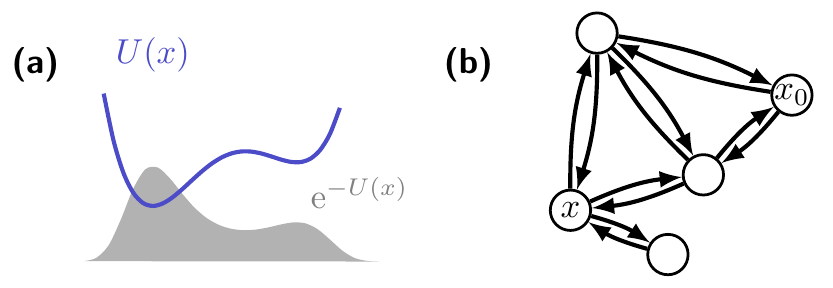}
 \caption{Schematic of (a) diffusive dynamics and (b) Markovian jump-process.}
 \label{fig:HartichGodec:relaxation}
\end{figure}
Having established that the full FPT statistics are required for a
correct physical description of reaction kinetics in the few-encounter
limit, we now present, on the hand of two illustrative examples, a canonical method to determine $\wp_a(t|x_0)$
from the corresponding relaxation spectrum. 

We consider two classes of processes, diffusion in effectively one-dimensional potentials and reversible Markovian jump-processes, in more
detail (see e.g. Fig.~\ref{fig:HartichGodec:relaxation}).
We call $x_t$ a relaxation process if, in contrast to the first
passage problem, the dynamics does not terminate upon reaching a threshold.
More precisely, for a diffusion process (see Fig.~\ref{fig:HartichGodec:relaxation}a) the probability density $P(x,t|x_0)$ to find a particle starting from $x_0$
at position $x$ at time $t$ satisfies the Fokker-Planck equation
\begin{equation}
\frac{\partial}{\partial t}P(x,t|x_0)=\hat{L}_\text{FP}P(x,t|x_0)\equiv
 \frac{\partial}{\partial x}D(x)\Big[ U'(x)+\frac{\partial}{\partial x}\Big]P(x,t|x_0),
 \label{eq:HartichGodec:FPE}
\end{equation}
where $U(x)$ is the potential $(U'\equiv\partial_x U)$
and $D(x)$ the diffusion landscape.
Relaxation dynamics conserves probability, i.e., $\int
P(x,t|x_0)\mathrm{d} x=1$ for all $t$, which is obtained either with
natural boundary condition or a ``reflecting barrier'', which would in
turn imply $[ U(x)-\partial_x]P(x,t|x_0)|_{x=a}=0$.
For jump-processes (see Fig.~\ref{fig:HartichGodec:relaxation}b) the probability to find the system at state $x_t=x$
if it started at $x_0$ obeys a master equation
\begin{equation}
\frac{\partial}{\partial t}P(x,t|x_0)=\hat{L}_\text{ME}P(x,t|x_0)\equiv\sum_{x'}
L_{xx'}P(x',t|x_0),
\label{eq:HartichGodec:ME}
\end{equation}
where $L_{xx'}$ is the transition rate from state $x'$ to $x$
if $x\neq x'$ and $L_{xx}=-\sum_{x'\neq x}L_{x'x}$ is the negative rate of leaving state
$x'$, guaranteeing conservation of probability $\sum_{x}L_{xx'}=0$, i.e., $\sum_{x} P(x,t|x')=1$ for all $t$ and $x'$.
Moreover, reversibility requires the rates to obey detailed balance $\ln(L_{xx'}/L_{x'x})=U(x')-U(x)$ \cite{kamp07}.
Both classes of reversible stochastic dynamics allow an expansion of
the operator $\hat{L}=\hat{L}_\text{FP},\hat{L}_\text{ME}$
in a real bi-orthogonal eigenbasis,
such that
\begin{equation}
 P(x,t|x_0)=\sum_{k\ge0}\psi_k^\mathrm{R}(x)\psi_k^\mathrm{L}(x_0)\mathrm{e}^{-\lambda_k t},
 \label{eq:HartichGodec:Pxx0}
\end{equation}
where $\lambda_k$  is the $k$th eigenvalue of operator $\hat L$
(with $0=\lambda_0<\lambda_1\le \lambda_2\le\ldots$),
and $\psi_k^\mathrm{R}$ ($\psi_k^\mathrm{L}$) are the corresponding right (left)  eigenvectors satisfying $\hat L\psi_k^\mathrm{R}=-\lambda_k\psi_k^\mathrm{R}$
and
$\psi_k^\mathrm{L}(x)=\mathcal{N}_k^{-1}\mathrm{e}^{ U(x)}\psi_k^\mathrm{R}(x)$.
The normalization for $\hat L=\hat L_\mathrm{FP}$ reads $\mathcal{N}_k=\int\mathrm{e}^{ U(x)}[\psi_k^\mathrm{R}(x)]^2\mathrm{d}x$,
whereas for $\hat L=\hat L_\mathrm{ME}$ the integral in $x$ becomes a sum.
Note that the zeroth eigenvector ($k=0$) is given by $\psi_0^\mathrm{R}(x)=\mathrm{e}^{- U(x)}$,
such that $\psi_0^\mathrm{R}(x)\psi_0^\mathrm{L}(x_0)=P^\text{eq}(x)$
is the Boltzmann distribution.

The terms $k>0$ in the sum of Eq.~\eqref{eq:HartichGodec:Pxx0}
relax to zero exponentially fast with rates $\lambda_k$,
and the corresponding eigenfunctions $\psi_k^\mathrm{R}(x)\psi_k^\mathrm{L}(x_0)$ quantify the redistribution of the probability mass.
For potential landscapes with $n$ energy basins (e.g., $n=2$ in left panel of Fig.~\ref{fig:HartichGodec:relaxation})
we generally expect at least one (or the last) gap at $\lambda_{n-1}\ll\lambda_{n}$ in the relaxation spectrum.

For any stationary Markov process the renewal theorem \cite{sieg51}
\begin{equation}
 P(a,t|x_0)=\int_0^tP(a,t-\tau|a)\wp_a(\tau|x_0) \mathrm{d}\tau,
 \label{eq:HartichGodec:renewal1}
\end{equation}
connects the propagator of relaxation dynamics to the FPT
density. It has the following intuitive interpretation:
if a particle starting from $x_0$ is found at position
$x_t=a$ at time $t$,
then it must have reached it for the first time before that time $\tau\le t$ and then returned to (or stayed at) $a$
in the remaining time interval $t-\tau$.
Laplace transforming Eq.~\eqref{eq:HartichGodec:renewal1}, where a convolution in the time domain becomes a product, translates Eq.~\eqref{eq:HartichGodec:renewal1} to
$\tilde P(a,s|x_0)=\tilde P(a,s|a)\tilde{\wp}_a(s|x_0)$,
i.e.,
\begin{equation}
 \tilde{\wp}_a(s|x_0)=\frac{\tilde P(a,s|x_0)}{\tilde P(a,s|a)}.
 \label{eq:HartichGodec:renewal2}
\end{equation}
Comparing Eq.~\eqref{eq:HartichGodec:renewal2} with
the first passage density \eqref{eq:HartichGodec:FPD_Laplace}
one can easily verify that poles of  the first passage time
distribution $\tilde\wp_a(s|x_0)$, which are located at the first passage rates $\mu_k=-s$,
are zeros of the diagonal of the propagator
$\tilde P(a,s|a)$ \cite{keil64}.

In Appendix~\ref{sec:HartichGodec:duality} we present
an explicit and exact duality relation that allows for an explicit
inversion of Eq.~(\ref{eq:HartichGodec:renewal2}) to the time
domain. Briefly, $\wp_a(t|x_0)$ is obtained in three steps: (i) the first step
is to realize that the first passage and relaxation time-scales
interlace, $\lambda_{k-1}\le\mu_k\le\lambda_k$, which is then utilized
in (ii) the second step to express all first passage rates $\{\mu_k\}$
in terms of series of determinants of almost triangular matrices
\eqref{eq:HartichGodec:muk}. (iii) The third an final step involves the Chauchy residue theorem to determine the first passage weights
$\{w_k\}$ from Eq.~\eqref{eq:HartichGodec:wk}, leading to
\begin{equation}
 w_k(x_0)=\frac{\sum_{l\ge0}(1-\lambda_l/\mu_k)^{-1}\psi^\mathrm{R}_l(a)\psi^\mathrm{L}_l(x_0)}{\sum_{l\ge0}(1-\lambda_l/\mu_k)^{-2}\psi^\mathrm{R}_l(a)\psi^\mathrm{L}_l(a)}.
 \label{weights}
\end{equation}
For the full details we refer the reader to Appendix
\ref{sec:HartichGodec:duality} or
Refs.~\refcite{hart18_arxiv,hart18a_arxiv}. In the following we apply
the duality to determine FPT densities of a simple four-state Markov
process and a diffusion in a rugged potential.

%
%
%
%
%

\subsection{Four state Markov jump process}
\label{sec:HartichGodec:example_discrete}
For illustratory purposes we consider a simple four state biased random walk as shown in the inset of
Fig.~\ref{fig:HartichGodec:pt_ill} with a transition matrix
\begin{equation}
 \mathbf{L}=
 \begin{pmatrix}
 -2 & 1 & 0 & 0 \\
 2 & -3 & 1 & 0 \\
 0 & 2 & -3 & 1 \\
 0 & 0 & 2 & -1 \\
 \end{pmatrix},
\end{equation}
whose eigenvalues are
$\{\lambda_0,\lambda_1,\lambda_2,\lambda_3\}=\{0,1,3,5\}$ and the
corresponding eigenvectors can be obtained in a straightforward
manner. We fix the initial and target state to $x_0=1$ and $a=4$,
respectively. The diagonal and off-diagonal relaxation propagators then have the simple forms
\begin{equation}
\begin{aligned}
 P(a,t|a)&=\frac{8}{15}+\frac{\mathrm{e}^{-t}}{4}+\frac{\mathrm{e}^{-3t}}{6}+\frac{\mathrm{e}^{-5t}}{20}\\
  P(a,t|x_0)&=\frac{8}{15}+\mathrm{e}^{-t}+\frac{2\mathrm{e}^{-3t}}{3}+\frac{\mathrm{e}^{-5t}}{5},
\end{aligned}
\label{eq:HartichGodec:Prob}
\end{equation}
whereas a similarly compact analytical formula for $\wp_a(t|x_0)$ cannot be found. In Appendix~\ref{sec:HartichGodec:4state_calc} we use the duality between relaxation and first passage processes to determine
$\{\mu_1,\mu_2,\mu_3\}\simeq \{0.657,2.529,4.814\}$
and
$\{w_1(x_0),w_2(x_0),w_3(x_0)\}\simeq\{1.565,-0.740,0.175\}$.
The resulting single-particle FPT probability density
is depicted with the solid line in the upper panel of
Fig.~\ref{fig:HartichGodec:pt_ill}. The dash-dotted line (here $n=2$)
in the lower panel depicts the corresponding single-particle survival probability $\mathcal{P}_a(t|x_0)$. We note that the short-time 
limit yields $\wp_a(t|x_0)=4 t^2+\mathcal{O}(t)^3$, which is arises from the two intermediate states between $x_0$ and $a$ (see model scheme from Fig.~\ref{fig:HartichGodec:pt_ill}).
The vanishing first passage density $\wp_a\to0$ (short-time limit) causes a strong narrowing of the many-particle first passage density $\wp_a^{(n)}(t|x_0)=n\wp_a(t|x_0)\mathcal{P}_a(t|x_0)^{n-1}$, since the survival probability
``pushes'' the probability mass to short times for increasing values of $n$ (see Fig.~\ref{fig:HartichGodec:pt_ill}).

\subsection{Diffusive exploration of a rugged energy landscape}
\label{sec:HartichGodec:example_rugged}
As a second example we analyze $\wp_a(t|x_0)$
for a diffusive barrier crossing in a rugged multi-well potential,
which is particularly relevant for protein folding 
and misfolding kinetics \cite{noe11,yu15,neup16,dee16} and biochemical association reactions \cite{schu81}.

\begin{figure}[t]
 \centering
 \includegraphics{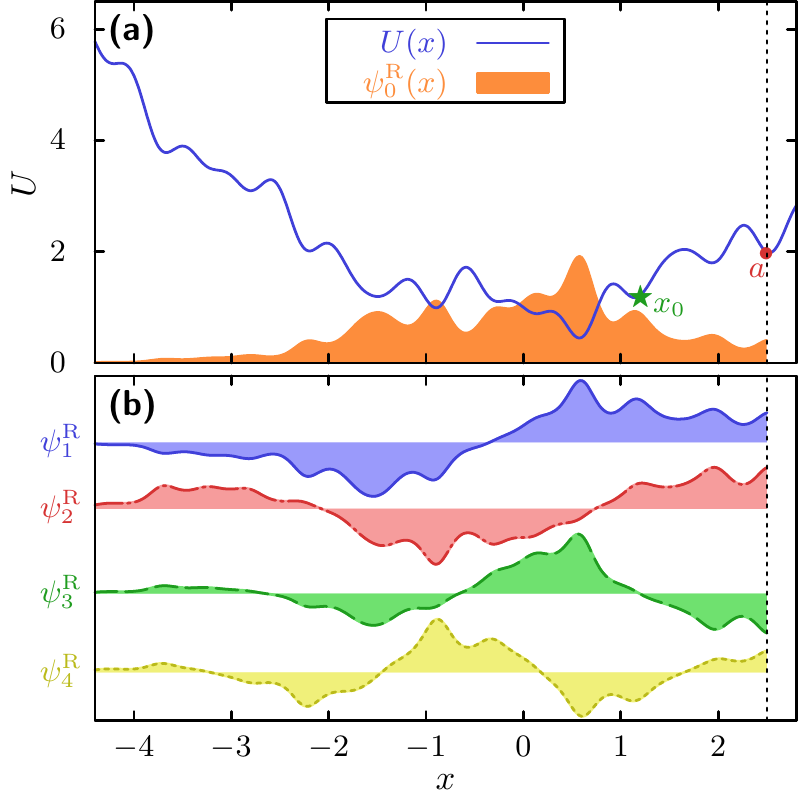}
 \caption{(a) Rugged potential landscape $U(x)$ from
   Eq.~(\ref{eq:HartichGodec:Karhunen}) and the corresponding Boltzmann measure with $(z_1,\ldots , z_7) =(-0.14,-1.04,0.77,-1.32,-0.61,-1.66,-2.67)$. (b) The first four
   excited right eigenfunctions corresponding to $U(x)$ with $D(x)=1$.}
 \label{fig:HartichGodec:rugged1}
\end{figure}

\begin{figure}[t]
 \centering
 \includegraphics{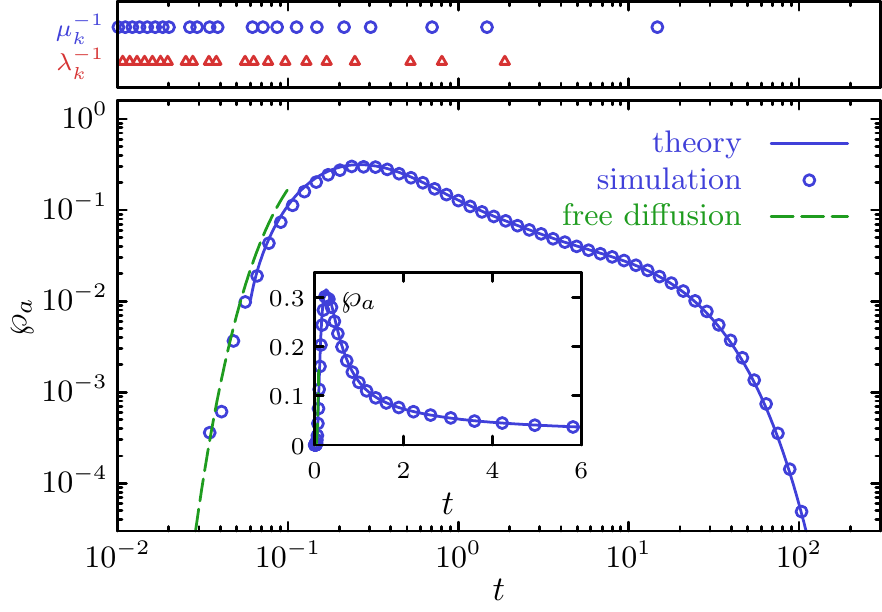}
 \caption{The FPT density for particle starting from $x_0=0.6$ and
 to $a=2.5$ within potential from Fig.~\ref{fig:HartichGodec:rugged1}. The inset
 shows the first passage time density on a linear scale.
 The upper panel superimposes the first passage time scales $\mu_k^{-1}$ (open circles)
 and the relaxation time scales $\lambda_k^{-1}$ (open triangles). The symbols are obtained using the theory outlined in
 Appendix \ref{sec:HartichGodec:duality}, and the symbols denote
 results of Brownian dynamics simulations of $10^6$ trajectories.}
 \label{fig:HartichGodec:rugged2}
\end{figure}

We generate a single rugged potential landscape as a sum of a harmonic potential and
a truncated Karhunen-Lo\`eve expansion of a Wiener process
\begin{equation}
 U(x)=\frac{x^2}{4}+\sum_{k=1}^Nz_k\frac{\sin[(2k-1)x]}{(2k-1)}.
 \label{eq:HartichGodec:Karhunen}
\end{equation}
We truncate the expansion at $N=7$ and sample $z_k$ from a normal
distribution. Once $\{z_k\}$ are determined, they are kept
fixed. In Fig.~\ref{fig:HartichGodec:rugged1}a
we depict $U(x)$ and its corresponding
equilibrium probability density $\psi_0^\mathrm{R}(x)\propto\mathrm{e}^{-U(x)}$. We numerically determine the first 45 eigenvalues
$\{\lambda_k\}$  and eigenfunctions $\{\psi_k^\mathrm{R}\}$ of the Fokker-Planck operator \eqref{eq:HartichGodec:FPE},
from which the first four excited relaxation eigenmodes are illustrated
in  Fig.~\ref{fig:HartichGodec:rugged1}b. The relaxation eigenfunctions determine the redistribution of probability
during  the approach to equilibrium.
Using the Newton's series of almost triangular matrices from
Eq.~\eqref{eq:HartichGodec:muk} we determine the first passage time-scales $\mu_k^{-1}$, which  interlace
with the relaxation time-scales \cite{hart18_arxiv,hart18a_arxiv} as illustrated in the upper panel of Fig.~\ref{fig:HartichGodec:rugged2}. Specifically, between any two consecutive first passage time scales
(blue circles) we find exactly one relaxation time-scale (red triangles)
and vice versa. In particular, the slowest first passage time-scale occurs on a longer time-scale
than the slowest relaxation time-scale.
This can be explained by the fact that the slowest first passage mode
requires \emph{all} trajectories to reach the target, whereas the slowest relaxation mode only reflects that most trajectories have reached the equilibrium distribution.

In the lower panel of Fig.~\ref{fig:HartichGodec:rugged2} we present results for the full FPT density using the analytical theory from Appendix~\ref{sec:HartichGodec:duality} (blue solid line),
together with results of extensive Brownian dynamics simulations of $10^6$ trajectories, which perfectly agree with the theory. The inset depicts $\wp_a(t|x_0)$ on linear scale. The short-time limit for a freely diffusing particle in form of a L\'evi-Smirnov density, also know as
Sparre Anderson result \cite{spar5X,gode16},
is shown as dashed green line (see also Refs.~\refcite{kamp93,beij03,meji11} for further discussions on the short-time limit). Intuitively, diffusion is faster
than advection on short time-scales ($\propto\sqrt{t}$ vs. $\propto t$ behavior), rendering the actual potential shape less relevant for $t\to0$.

\section{Concluding perspectives}
\label{sec:HartichGodec:conclusion}
The mean and higher moments of the FPT in a single-particle setting
were shown to be inherently insufficient for characterizing many-particle
FPT kinetics within the few-encounter limit. To correctly describe few-encounter kinetics
one has to go beyond a description limited to FPT moments and determine the full FPT distribution.
It was shown how to achieve this utilizing a duality relation between relaxation and first passage
process \cite{hart18_arxiv,hart18a_arxiv} outlined in Appendix \ref{sec:HartichGodec:duality}.
The method is applicable to a broad class of reversible Markov dynamics that includes discrete Markovian jump-processes in any dimension and
Markovian diffusion in effectively one-dimensional potential landscapes.

The duality relation can in fact be considered as an analytical algorithmic tool
for determining FPT distributions, which was demonstrated on hand of a simple four state model in full detail.
The analysis of the $n$-particle FPT distribution revealed a reduced mean FPT and a
canonical narrowing of the FPT distribution in the few-encounter limit
as the number of particle increases. This narrowing arises due to a
combination of the
short-time cutoff in the FPT density ($\wp_a\to0$ for $t\to 0$)
and an inherent many-particle speed-up, which together render the
$n$-particle kinetics deterministic in the limit $n\to\infty$. In the
case of a diffusive exploration of (rugged) energy landscapes the
short-time behavior is dominated by free diffusion,
rendering the shape of the potential essentially  irrelevant\cite{kamp93,beij03}.

It will be interesting to extend the applications of the theory outlined in Appendix
\ref{sec:HartichGodec:duality} and to explore the physical
consequences of few-encounter kinetics also in narrow escape problems
\cite{sing06,schu07,rein09,pill10,isaa16,greb17} and diffusion on
higher-dimensional graphs. Extending the work to irreversible dynamics will be challenging, whereas long-time asymptotics are still accessible \cite{gode16, hart18a_arxiv}.


\appendix
\section{Duality between relaxation and first passage processes}
\label{sec:HartichGodec:duality}
\subsection{General case}
In this appendix we review the duality relation from
Refs.~\refcite{hart18_arxiv,hart18a_arxiv} that allows us to determine analytically the spectral 
representation of the FPT density in
Eq.~\eqref{eq:HartichGodec:spectral_def} from the propagator in Eq.~\eqref{eq:HartichGodec:Pxx0} in three steps.

The first step is to realize that the relaxation time-scales $\{\lambda_k^{-1}\}$ and first passage times-scales $\mu_k^{-1}$ interlace \cite{hart18_arxiv,hart18a_arxiv}
\begin{equation}
 \lambda_{k-1}\le \mu_k\le \lambda_k.
\end{equation}
We note that this interlacing of time-scales can be related to Chauchy's interlacing theorem
for real symmetric matrices \cite{gron08}. The interlacing has also been
demonstrated for simple one-dimensional processes \cite{keil64}.

The second step is based on an explicit Newton iteration
that allows, after to some rather involved
algebra \cite{hart18_arxiv, hart18a_arxiv},
to exactly express the
first passage
rates $\mu_k$ as a series of determinants of almost triangular matrices $\boldsymbol{\mathcal{A}}_n(k)$
\begin{equation}
 \mu_k=\bar\mu_k+\sum_{n=1}^\infty f_0(k)^nf_1(k)^{1-2n}\det\boldsymbol{\mathcal{A}}_n(k),
 \label{eq:HartichGodec:muk}
\end{equation}
where $\bar\mu_k\equiv(\lambda_k+\lambda_{k-1})/2$,
$f_n(k)=\partial_s^n F(k^*,s)|_{s=-\bar{\mu}_k}$ with 
\begin{equation}
F(k,s)=(s+\lambda_k)\tilde P(a,s|a)
\label{eq:HartichGodec:FK}
\end{equation}
and the index function
\begin{equation}
 k^*\equiv k^*(k)=
 \begin{cases}
  k&\text{if $F(k,-\bar\mu_k)<0$},\\
  k-1&\text{else}
 \end{cases}
 \label{eq:HartichGodec:k*}
\end{equation}
that guarantees $f_0(k)$ to be negative, and we used the almost triangular $(n-1)\times(n-1)$ matrices with elements
\begin{equation}
 \mathcal{A}_n^{i,j}(k)=\frac{f_{i-j+2}(k)\Theta(i-j+1)}{(i-j+2)!}
 \begin{cases}
  i+j-1&\text{if $j=1$,}\\
  n(i-j+1)+j-1&\text{if $j>1$,}
 \end{cases}
 \label{eq:HartichGodec:Aij}
\end{equation}
where $\Theta$ is the discrete Heaviside step function ($\Theta(l)=1$ if $l\ge0$) and $\det \boldsymbol{\mathcal{A}}_1(k)=1$.
Moreover, we have explicitly \cite{hart18_arxiv,hart18a_arxiv}
\begin{equation}
\begin{aligned}
f_0(k)&=\psi^\mathrm{L}_{k^*}(a)\psi_{k^*}^\mathrm{R}(a)+\sum_{l|l\neq k^*}\psi^\mathrm{L}_l(a)\psi_l^\mathrm{R}(a)\frac{(\bar\mu_k-\lambda_{k^*})}{(\bar\mu_k-\lambda_l)},\\
f_{n\ge 1}(k)&=n!\sum_{l|l\neq k^*}\psi^\mathrm{L}_l(a)\psi_l^\mathrm{R}(a)\frac{(\lambda_l-\lambda_{k^*})}{(\bar\mu_k-\lambda_l)^{n+1}}.
\label{eq:HartichGodec:f_n(k)}
\end{aligned}
\end{equation}

The third step is a straightforward application of the
residue theorem, delivering the first passage
weights
\begin{equation}
 w_k(x_0)=\frac{ \tilde P (a,s|x_0)}{\mu_k\partial_s\tilde P (a,s|a)}\bigg|_{s=-\mu_k}=\frac{\sum_{l\ge0}(1-\lambda_l/\mu_k)^{-1}\psi^\mathrm{R}_l(a)\psi^\mathrm{L}_l(x_0)}{\sum_{l\ge0}(1-\lambda_l/\mu_k)^{-2}\psi^\mathrm{R}_l(a)\psi^\mathrm{L}_l(a)},
 \label{eq:HartichGodec:wk}
\end{equation}
where $\tilde P (a,s|x_0)$ is the Laplace transform of Eq.~\eqref{eq:HartichGodec:Prob}.
We note that Eq.~\eqref{eq:HartichGodec:muk} and \eqref{eq:HartichGodec:wk} are exact relations that fully characterize
the first passage kinetics.
 
\subsection{Four state model}
\label{sec:HartichGodec:4state_calc}
We now evaluate $\wp_a(t|x_0)$ for the model from Sec.~\ref{sec:HartichGodec:example_discrete} step by step.
First, the Laplace transform of the first line of
Eq.~\eqref{eq:HartichGodec:Prob}, $\tilde P(a,s|a)$, is inserted into Eqs.~\eqref{eq:HartichGodec:FK} and \eqref{eq:HartichGodec:k*} giving $k^*=k$ for $k=1,2,3$.
Second, Eq.~\eqref{eq:HartichGodec:f_n(k)}
yields
\begin{equation}
 \begin{pmatrix}
  f_0(1)\\
  f_0(2)\\
  f_0(3)
 \end{pmatrix}
 =
  \begin{pmatrix}
  -\frac{11}{45}\\
  -\frac{1}{3}\\
  -\frac{1}{3}
 \end{pmatrix},
 \qquad
  \begin{pmatrix}
  \frac{f_n(1)}{n!}\\
  \frac{f_n(2)}{n!}\\
  \frac{f_n(3)}{n!}
 \end{pmatrix}
 =
 \begin{pmatrix}
  \frac{1}{3}(-\frac{2}{5})^{n+1}+\frac{1}{5}(-\frac{2}{9})^{n+1}-\frac{2^{n+4}}{15}\\
  \frac{1}{10}[(-3)^{-n-1}-2^{3-n}-5]\\
  \frac{1}{3} [-2^{1-2 n}-3^{-n}-1]
 \end{pmatrix}
\end{equation}
with $n>0$.
Note that $k^*$ is chosen to guarantee the negativity of $f_0(k)$. Third, inserting the $f_n(k)/n!$ into the almost triangular matrix
Eq.~\eqref{eq:HartichGodec:Aij} and evaluating
the Newton's series \eqref{eq:HartichGodec:muk} yields the exact first
passage time-scales, which numerically are given by $\{\mu_1,\mu_2,\mu_3\}\simeq \{0.657,2.529,4.814\}$.
Finally, the weights from Eq.~\eqref{eq:HartichGodec:wk}
yield
$\{w_1(x_0),w_2(x_0),w_3(x_0)\}\simeq\{1.565,-0.740,0.175\}$, which fully determines $\wp_a(t|x_0)$.

\section*{Acknowledgements}
The financial support from the German Research
Foundation (DFG) through the Emmy Noether Program
``GO 2762/1-1'' (to AG) is gratefully acknowledged.

%
%

\end{document}